\documentclass[prd,aps,secnumarabic,showpacs]{revtex4}
\usepackage{amsmath}
\usepackage{amsfonts}
\usepackage{amssymb}
\usepackage{graphicx,color}
\def\sb{\mbox{\rule[-4pt]{0pt}{14pt}}}
\def\tb{\mbox{\rule{12pt}{0pt}}}

\def\ga{\gamma}

\def\ep{\varepsilon}

\def\part{\partial}

\begin{document}

\title{Dalitz Decay $H\rightarrow\,f\bar{f}\gamma$ as a Background for $H\rightarrow\gamma\,\gamma$}

\author{Duane A. Dicus$^a$\footnote{Email address: dicus@physics.utexas.edu}
and Wayne W. Repko$^b$\footnote{Email address: repko@pa.msu.edu}}

\affiliation{
$^a$Department of Physics and Center for Particles and Fields,
University of Texas, Austin, TX 78712, USA \\
$^b$Department of Physics and Astronomy, Michigan State University, East Lansing, Michigan 48824, USA}
%

\date{\today}

\begin{abstract}
The Dalitz decay $H\rightarrow\,f\bar{f}\gamma$ is calculated for very small $f\bar{f}$ invariant masses where the $f\bar{f}$ pair could be mistaken for a photon in analysis of $H\rightarrow\gamma\gamma$ decays. Using the ATLAS cuts and the full Dalitz decay amplitude, we estimate this fraction to be 7.06\%.
\end{abstract}


\maketitle

The identification of the Higgs boson candidate at the LHC \cite{Higgs1,Higgs2} relies heavily on its decays into vector bosons, $H\to\ga\ga$, $H\to\ga Z$ and $H\to ZZ^*$. Here, we focus on the sources of background to the di-photon decay mode from the Dalitz decay $H\to f\bar{f}\ga$. There have been several recent calculations of the width for Higgs decay into $\ell^{+}\ell^{-}\gamma$ where the $\ell$ are leptons 
\cite{DR,CQZ,GP,SCG,DKW}. In particular, in Ref.\,\cite{DKW}, we looked at large $\ell\bar{\ell}$ invariant masses to compare with $H\rightarrow\gamma\,Z, Z\rightarrow\,\ell\bar{\ell}$. Here we want to consider $H\rightarrow\,f\bar{f}\gamma$ for $f\,=\,e,\mu,\tau,u,d,s,c$ with very small $f\bar{f}$ invariant masses and ask for the probability that the $f\bar{f}$ will be mistaken for a photon and contaminate the measurement of the partial width $H\rightarrow\gamma\gamma$. At least one earlier paper, Ref.\,\cite{Strn}, has considered this question and estimated the fraction of the width of $H\rightarrow\gamma\gamma$ that is really $H\rightarrow\,f\bar{f}\gamma$ in disguise.  We will compare with their results.   

We use the experimental cuts of the ATLAS group. In the ATLAS detector, a fermion-antifermion pair can be mistaken for a photon if $m_{f\bar{f}}\,<\,10$ GeV. In addition there are visibility cuts on the photon and on $f\bar{f}$. In terms of transverse momentum and rapidity these are
\begin{subequations}
\begin{eqnarray}
{\rm Min}(pT_{f\bar{f}},\,pT_{\gamma})\,&>&\,0.25\times 125\,{\rm GeV}\,,  \\ 
{\rm Max}(pT_{f\bar{f}},\,pT_{\gamma})\,&>&\,0.35\times 125\,{\rm GeV}\,,  \\ 
|\eta_{\gamma}|\,&<&\,2.37\,,  \\ 
|\eta_{f\bar{f}}|\,&<&\,2.37\,.
\end{eqnarray}
\end{subequations}
For $H\to\ga\ga$, take $f\bar{f}$ to be the second photon.

First, we calculate the fraction $f$ that satisfies these cuts as
\begin{equation}\label{cut}
f\,=\,\frac{\int^1_0dx_1\int^1_0\,dx_2\,g(x_1,M_H)g(x_2,M_H)\delta(\hat{s}-M_H^2)\int^1_{-1}\,dz}
           {\int^1_0dx_1\int^1_0\,dx_2\,g(x_1,M_H)g(x_2,M_H)\delta(\hat{s}-M_H^2)\,2}\,,
\end{equation}
where the $g(x,M_H)$ are gluon distribution functions that weight the average. In the numerator we use $x_1$ and $x_2$ to define a $p_1^{\mu}$ and $p_2^{\mu}$
\begin{subequations}
\begin{eqnarray}
p_1^{\mu}\,&=&\,(x_1\,E_b,0,0,x_1\,E_b)\,\,,\\ 
p_2^{\mu}\,&=&\,(x_2\,E_b,0,0,-x_2\,E_b)\,,
\end{eqnarray}
\end{subequations}
where $E_b$ is the beam energy, and the momentum of the Higgs is $p_1+p_2$. The momenta of the photons in the Higgs rest frame are given by $z$
\begin{subequations}
\begin{eqnarray}
k_1^{\mu}\,&=&\,\frac{M_H}{2}(1,\sqrt{1-z^2},0,z)\,,\\
k_2^{\mu}\,&=&\,\frac{M_H}{2}(1,-\sqrt{1-z^2},0,-z)\,.
\end{eqnarray}
\end{subequations}
$k_1$ and $k_2$ are boosted into the frame $p_1+p_2$ and if they fail the cuts $f$ is set to zero. The CTEQ6L1 parton distribution functions \cite{CTEQ} were used for the gluons. Because the distribution functions occur in both the numerator and denominator, Eq.\,(\ref{cut}) is not sensitive to their precise form.

For the narrow range of values of $M_H$ of interest the result for $f$ is almost independent of $M_H$: $f_4\,=\,0.610$ for $E_b\,=\,4$ TeV and $f_7\,=\,0.559$ for $E_b\,=\,7$ TeV. The result, $f\,\Gamma_{H\rightarrow\gamma\gamma}$, depends on the beam energy and the Higgs mass. It is shown in Table I as $\Gamma_4$ and $\Gamma_7$ for beam energies of $4$ or $7$ TeV.

\begin{table}[h]
\begin{tabular}{|c|c|c|c|} \hline
\,\,\,$M_H$\,(GeV)\,\,\,\,&\,\,\,\,$\Gamma_{H\rightarrow\gamma\gamma}$ (keV)\,\,\,\,&\,\,\,$\Gamma_{4}$\,(keV)\,\,\,&\,\,\,$\Gamma_{7}$\,(keV)\,\,\\ \hline
124 & 10.10 & 6.15 & 5.63  \\
125 & 10.45 & 6.37 & 5.83  \\
126 & 10.81 & 6.60 & 6.04  \\
127 & 11.19 & 6.83 & 6.25  \\ 
\hline
\end{tabular}
\caption [Table I] {The full width $\Gamma_{H\rightarrow\gamma\gamma}$ and the widths $\Gamma_4$ and $\Gamma_7$ after cuts for beam energies of $4$ TeV and $7$ TeV are shown.}
\end{table}

\begin{figure}[h]\centering
\includegraphics[height=0.9in]{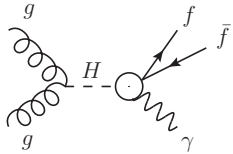}
\caption{The gluon fusion process for the Higgs production with one photon decaying into an $f\bar{f}$ pair is shown.} \label{Hffg}
\end{figure}
For $H\rightarrow\,f\bar{f}\gamma$, illustrated in Fig.\,\ref{Hffg},  we proceed as in Ref.\,\cite{DKW} using the cuts above and an $f\bar{f}$ invariant mass less than $10$ GeV. The width in the Higgs rest frame is
\begin{equation}\label{G0}
\Gamma_0\,=\,\frac{\pi}{32}\frac{1}{(2\pi)^5}\frac{1}{2M_H}\int^{2\pi}_0\, d\phi\,\int^1_{-1}\,dz\,\int^{M_H^2}_{4m^2}\,ds\,\int^1_{-1}\,dz' \sqrt{1-\frac{4m^2}{s}}\left(1-\frac{s}{M_H^2}\right)|M|^2
\end{equation}
where $m$ is the fermion mass, $s$ is the invariant mass of the fermions, and $|M|^2$ is the square of the matrix element as given in Ref.\,\cite{ABDR}. The variables $z$ and $\phi$ define the momenta of the individual fermions, $p^{\mu}$ and $p'^{\mu}$, in their center of mass while $z'$ is used to define the momentum of the photon, $k^{\mu}$, and the momentum of the fermion pair, $Q^{\mu}$, in the Higgs rest frame. To find the width in the boosted frame, where we can apply the cuts given above, we boost $k^{\mu}$ and $Q^{\mu}$ into the lab frame given by $p_1^{\mu}\,+\,p_2^{\mu}$ and then boost $p^{\mu}$ and $p'^{\mu}$ into the frame now given by the boosted $Q^{\mu}$. As in Eq.(\ref{cut}), we average over the possible lab frames
\begin{equation}
\Gamma\,=\,\frac{1}{N}\int^1_0dx_1\,\int^1_0dx_2\,\,g(x_1,M_H)g(x_2,M_H)\delta(\hat{s}-M_H^2)\Gamma_0
\end{equation}
where $\hat{s}\,=(p_1+p_2)^2\,=\,4E_b^2x_1x_2$. $N$ is the same integral without the $\Gamma_0$ and is included to normalize the average. For each value of $x_1$ and $x_2$ we check the cuts and put $\Gamma$ to zero if they are not respected.

As it turns out, the results, when written as a fraction of $\Gamma_{4}$ or $\Gamma_{7}$, are independent of $M_H$ or beam energy. The contributions from each final state are listed in Table II as $\ep_{f\bar{f}}$. The total fraction of the measured $H\rightarrow\gamma\gamma$ width that is erroneously due to $H\rightarrow\,f\bar{f}\gamma$ is $7.06\%$. This problem was considered in Ref.\,\cite{Strn} but without doing a full calculation of the width or applying cuts.  Nevertheless their results are a good estimate -- our numbers are somewhat smaller, particularly for $\mu$, $\tau$, $c$, and $b$.  Their total correction is $10.22\%$. 

In summary, we have used the exact $H\to f\bar{f}\ga$ amplitudes \cite{ABDR} together with the cuts adopted by the ATLAS collaboration to provide a more reliable estimate of the contamination of the measured $H\to\ga\ga$ width due to the background from the Dalitz decays with low invariant mass $f\bar{f}$ pairs.

{\bf Acknowledgements} -- It is our pleasure to thank the ATLAS group whose questions led us to do this work, F. Bernlochner, D. Gillberg, J. Saxon, and K. Tackmann. D.A.D. was supported in part by the U.S.Department of Energy under Award No. DE-FG02-12ER41830. W.W.R. was supported in part by the National Science Foundation under Grant No. PHY 1068020.
\begin{table}[h]\centering
\begin{tabular}{|c|c|} \hline
\sb\tb $f\bar{f}$ \tb   & \tb $\ep_{f\bar{f}}$ \tb\\ \hline
$e\bar{e}$        & 2.90 \\
$\mu\bar{\mu}$    & 1.14 \\
$\tau\bar{\tau}$  & 0.23 \\
$u\bar{u}$        & 1.71 \\
$d\bar{d}$        & 0.42 \\
$s\bar{s}$        & 0.21 \\
$c\bar{c}$        & 0.45 \\ \hline
Total             & 7.06 \\ \hline
\end{tabular}
\caption [Table II ] {The second column is the fraction of the $H\to\ga\ga$ width the can be attributed to $H\to f\bar{f}\ga$ decays given as a percentage. For example, $2.90\%$ of the $H\to\ga\ga$ decays are actually $H\to e\bar{e}\ga$. The contribution of $b\bar{b}$ is $0.002\%$. For $u$ and $d$ we took the mass to be the half the pion mass.  For $s$ we used the kaon mass.}
\end{table}

\end{document}